%% file: main.tex
\documentclass{hotnets22}

\input{src/preamble.tex}
\title{Superflows: A New Tool for Forensic Network Flow Analysis}
\author{
  Michael Collins$^1$ \and
  Jyotirmoy V. Deshmukh$^1$ \and
  Dristi Dinesh$^1$ \and
  Mukund Raghothaman$^1$ \and
  Srivatsan Ravi$^1$ \and
  Yuan Xia$^1$\\
  \centering $^1$University of Southern California
}

\begin{document}

\maketitle

\input{src/abstract.tex} \mclearpage
\input{src/intro.tex} \mclearpage
\input{src/related.tex} \mclearpage
\input{src/sf.tex} \mclearpage
\input{src/studies.tex} \mclearpage
\input{src/future.tex}

\bibliographystyle{abbrv}
\bibliography{src/references}


\end{document}

%% file: src/preamble.tex
\usepackage{times}
\usepackage{hyperref}
\usepackage{titlesec}
\usepackage{cleveref}
\hypersetup{pdfstartview=FitH,pdfpagelayout=SinglePage}

\usepackage{amsmath, amssymb, amsthm}
\usepackage{comment}
\usepackage{graphicx} 
\usepackage{microtype}
\usepackage{multicol}
\usepackage{paralist}

\newtheorem{thm}{Theorem}
\newtheorem{claim}[thm]{Claim}

\newtheorem{example}[thm]{Example}

\newcommand{\eg}{{\em e.g.}}
\newcommand{\ie}{{\em i.e.}}

\newcommand{\Flow}{{\operatorname{Flow}}}
\newcommand{\Bool}{{\operatorname{Bool}}}
\newcommand{\true}{{\operatorname{true}}}
\newcommand{\false}{{\operatorname{false}}}
\newcommand{\attr}{{\operatorname{attr}}}
\newcommand{\srcip}{{\operatorname{srcip}}}
\newcommand{\dstip}{{\operatorname{dstip}}}
\newcommand{\dstport}{{\operatorname{dstport}}}
\newcommand{\timestart}{{t_{\text{start}}}}
\newcommand{\timeend}{{t_{\text{end}}}}
\newcommand{\numbytes}{{\operatorname{\#bytes}}}
\newcommand{\numpackets}{{\operatorname{\#packets}}}
\newcommand{\decomp}{{\operatorname{decomp}}}

\newcommand{\union}{\mathbin{\cup}}

\newcommand{\mclearpage}{}

%% file: src/abstract.tex
\begin{abstract}
Network security analysts gather data from diverse sources, from high-level summaries of network flow and traffic volumes to low-level details such as service logs from servers and the contents of individual packets.  They validate and check this data against traffic patterns and historical indicators of compromise. Based on the results of this analysis, a decision is made to either automatically manage the traffic or report it to an analyst for further investigation. Unfortunately, due rapidly increasing traffic volumes, there are far more events to check than operational teams can handle for effective forensic analysis. However, just as packets are grouped into flows that share a commonality, we argue that a high-level construct for grouping network flows into a set a flows that share a hypothesis is needed to significantly improve the quality of operational network response by increasing Events Per Analysts Hour (EPAH). 


In this paper, we propose a formalism for describing a \emph{superflow} construct, which we characterize as an aggregation of one or more flows based on an analyst-specific hypothesis about traffic behavior. We demonstrate simple superflow constructions and representations, and perform a case study to explain how the formalism can be used to reduce the volume of data for forensic analysis.
\end{abstract}

%% file: src/intro.tex
\section{Introduction}
\label{sec:intro}

We propose a new form of NetFlow-like constructs for security
analysis, which we call {\em SuperFlows}.  SuperFlows group multiple
individual flows together around a common hypothesis, such as that all
the flows represent a single webpage fetch, a scan or a DGA exploit.

Superflows are built out of flows, which were originally developed for
traffic measurement~\cite{claffy94}.  Security analysts adopted flow
~\cite{fullmer00}, and developed tools analyzing flow and adding new
flow attributes for collection.  For security analysis, flow provides
enormous "bang for the buck" -- flows provide a compact summary of the
most important information about a session.  This compact information
is critical -- flows enable analysts to quickly examine large sets of
traffic and infer potentially hostile behavior.  The amount of
information an analyst needs to examine for a particular flow, in
terms of the footprint on disk, is far smaller for a NetFlow -- easily
three or more orders of magnitude, then for a corresponding full pcap
session.

We contend that there are now two classes of NetFlow analysis with
different data collection needs: {\em traffic} and {\em security}.
Traffic analysis, which is focused on billing and continuity needs,
use sampled NetFlow to understand the normal course of operations;
this has led to new statistical summary techniques, notably
sketches~\cite{liu16} which are sampling based, at the point of
collection, and based on soft real-time constraints.  Forensic
analysis requires the ability to reconstruct rare events, leading to a
specific {\em forensic} need for unsampled
Netflow~\cite{gall20,gall23,thomas14}, requiring new summary
constructs to reduce the data footprint while still providing evidence
for every network session.


SuperFlows are motivated by the need for traffic summaries describing
modern network traffic.  The characteristic traffic of the early 1990's
was the telnet session -- a TCP moderated, long-lived connection where
a single user communicated from a single client on a single host to a
single server.  The characteristic traffic of the modern era is the
webpage -- an assemblage of files fetched from dozens of servers, many
of whom are geographically distributed clones.  This characteristic
single client/multiple server behavior also defines many other
behaviors, from simple client-server interaction (because DNS TTL's
have dropped to values so low that name lookups are continuous), to
scanning, to torrenting, to serverless architectures.

We envision superflows as a new class of summaries that {\em supplement} raw flow data; when presented with a set of traffic for analysis, the analysis may be presented initially with multiple superflows that group the constituent flows together through their hypotheses of what the traffic represents.  For example, instead of seeing a dozen individual HTTPS flows, the analyst may see a single superflow marked "webpage fetch: news website.com", along with equivalent high-value summary information. If the analyst needs to examine that phenomenon in more depth, they can then pull up the individual flows based on guidance provided by the superflow.  For this approach to be effective, the superflow must be compact and the hypothesis guiding its creation must be clear, unambiguous and easily communicated to other users.  

\noindent\textbf{Contributions and roadmap}
The motivation for developing SuperFlows is to create a \emph{universal} formalism for expressing such hypotheses over
flow sequences. Furthermore, we provide an algorithm
for efficiently identifying the
maximal subsets of a flow sequence that satisfy large classes of superflow hypotheses. We introduce two case studies for demonstrating the superflow construct: (i) analysing scan data from an institutions dark spaces, (ii) performing a modern webpage analysis for understanding the interaction of protocols in multiple flows to produce a single web page.


The paper's contributions are structured as follows: we provide a historical motivation for the superflow construct and prior attempts at traffic analysis of aggregated flows (\cref{sec:related}). We introduce a language based on relational logic to express superflow hypotheses and how we can constructively identify the
maximal subsets of a flow sequence that satisfy the hypothesis (\cref{sec:sf}).
We present the efficacy of the formalism to improve the quality of operational network response through two case studies (\cref{sec:studies}. We conclude with an important discussion of the missing dimensions in our formalism: how the \emph{vantage} points for data collection will need to be incorporated into our formalism as well the role that \emph{confounders} (like NAT boxes) will need to be expressible within our framework for widespread applicability of the superflow construct.

%% file: src/related.tex
\section{Motivation and Related Work}
\label{sec:related}
While originally developed for traffic reporting, NetFlow rapidly developed into a forensic tool with the development of analysis packages such as Fullmer and Romig's Flow-Tools~\cite{fullmer00} and the CERT's SiLK Suite~\cite{gates04}.  These tools effectively mapped the relational calculus to NetFlow format flat files and, in the course of developing security analysis identified additional fields needed during flow collection.  This work partially informed the development of the IPFIX~\cite{rfc7011} standard, whose reference implementation, YAF~\cite{inacio10}, was developed in-house at the CERT, IPFIX includes fields for forensics, notably initial packet flags for TCP.  

Outside of this initial work, we have seen two distinct classes of traffic summarization which differ over the role of sampling and estimation.  We view the primary difference between these two classes as a comfort with statistical estimation.  The first class is focused on traffic summarization techniques and relies on sampling and estimation heavily, the second class is focused on forensic reconstruction and operational security, which increasingly views unsampled NetFlow as mission critical~\cite{cisco22,daniels20}.  

The first class of traffic analyses are {\em comfortable with sampling}, and are focused on developing highly-efficient soft-realtime summaries, primarily using streaming approaches.  The largest group of these techniques are based around various {\em sketch-based} algorithms~\cite{liu16,namkung22}, are focused on highly efficient streaming estimates of specific traffic characteristics (\eg, Entropy~\cite{clifford13}, traffic changes~\cite{krishnamurthy03}, heavy hitters~\cite{vibhaalakshmi17}).

The second class consists of techniques to identify and behaviorally summarize different traffic classes~\cite{paredes10}.  These techniques include systems which create constructs from traffic data, including the SiLK Set and Bag~\cite{mchugh04}, which group together arbitrary collections of IP addresses, and the FlowTuple~\cite{alcock21}, developed as part of CAIDA's Corsaro toolkit~\cite{corsaro}.  Other work involves techniques for identifying specific traffic classes such as botnets~\cite{zhang14,yen10}, scanning~\cite{collins08}, or peer-to-peer filesharing~\cite{collins06}.  \emph{These approaches represent different ways of identifying traffic phenomena, but each one is a separate detector; superflows are intended to unify these different cases into a common data reduction format to improve analyst workflow.}

%% file: src/sf.tex
\section{Superflow Decompositions}
\label{sec:sf}

Superflows provide a mechanism for security analysts to group NetFlow records by means of
user-provided hypotheses.
Formally, a \emph{superflow hypothesis} is a predicate over sets of flows, $h : 2^\Flow \to \Bool$.
For example, a set of flows $F = \{ f_1, f_2, \dots, f_k \}$ may arise from a scan if they all share
the same source IP address, attempt to reach hosts within the same subnet, occur within a short time
of each other, and probe a sufficiently large set of destinations. The analyst may operationalize
this by declaring that the set of flow records $F$ satisfies the superflow hypothesis
$h_{\text{scan}}$ when:
\begin{alignat}{1}
  h_{\text{scan}}(F) & = \forall f, f' \in F, \srcip(f) = \srcip(f') \land{} \nonumber \\
                     & \qquad \qquad \dstip(f) \sim \texttt{192.168.1.*} \land{} \nonumber \\
                     & \qquad \qquad \timestart(f') - \timestart(f) \leq 10 \text{ s} \land{} \nonumber \\
                     & \qquad \qquad |\{ \dstip(f) \mid f \in F \}| \geq c,
  \label{eq:sf:scan}
\end{alignat}
where $c$ is an analyst-supplied threshold value.

Upon examining a stream of NetFlow records, the analyst might mentally group flows that appear to
arise from a common underlying event such as $h_{\text{scan}}$. They may then proceed either to
further analyze individual hypothesized scan events, or examine flows which do not appear to match
the scan hypothesis.

Given a superflow hypothesis $h$, a \emph{superflow decomposition} $\decomp_h(F)$ of a set of flows
$F = \{ f_1, f_2, \dots, f_n \}$ is a partition of $F$ into disjoint subsets,
\[
  F = F_1 \union F_2 \union \dots \union F_k \union F_{\text{rest}},
\]
such that $h(F_i) = \true$ for $i = 1, 2, \dots, k$. Naturally, the analyst might be interested in
\emph{maximally} decomposing $F$, so they may additionally stipulate that:
\begin{inparaenum}[(\itshape a\upshape)]
\item $h(F_r') = \false$, and that
\item $h(F_i \union F_r') = \false$, for all subsets $F_r' \subseteq F_{\text{rest}}$, and for $i =
  1, 2, \dots, k$.
\end{inparaenum}
We call each of the partitions, $F_i$, for $i = 1, 2, \dots, k$ a \emph{superflow}.

As we will observe in Section~\ref{sec:studies}, grouping flows in this manner massively shrinks
both the set of observed events and the anomalous flows needing further investigation. We will now
describe a simple language for analysts to describe rich superflow hypotheses, and an efficient
algorithm to perform maximal superflow decompositions.


\subsection{A Language for Superflow Hypotheses}
\label{sub:sf:lang}

\paragraph{Attributes and predicates over flows.}

Unlike traditional flow grouping constructs such as \texttt{rwgroup} which is included as part of
the SiLK analysis suite, superflows allow us to group flows based on complex properties. Our
language for superflow hypotheses is inspired by relational logic, similar to that used in modeling
systems such as Alloy~\cite{Alloy}. We begin by fixing a set of flow attributes:
\[\begin{array}{rcl}
  \attr & ::=  & \srcip \mid \dstip \\
        & \mid & \timestart \mid \timeend \\
        & \mid & \numbytes \mid \numpackets \\
        & \mid & \cdots
\end{array}\]
Each of these attributes is a function which returns the corresponding property of the flow in
question, $\attr(f)$. A natural choice for these attributes are the properties exported as part of
the IPFIX record~\cite{}.
We also fix a set of atomic predicates over flows, $p$, $q$, \ldots, of varying arities. Examples
include unary predicates such as $\dstip(f) \sim \texttt{192.168.1.*}$ and binary predicates such as
$\srcip(f) = \srcip(f')$ and $\timestart(f) - \timestart(f') \leq 10 \text{ s}$.

\paragraph{Relational constraints over multiple flows.}

Superflow hypotheses may now be constructed as closed first-order logical formulas with cardinality
constraints:
\[\begin{array}{rcl}
  h & ::=  & \forall f, h \mid \exists f, h \\
    & \mid & h_1 \land h_2 \mid h_1 \lor h_2 \mid \lnot h \\
    & \mid & p(f_1, f_2, \dots, f_k) \\
    & \mid & |\{ \attr(f) \mid p(f) \}| \bowtie c, \text{ for } \bowtie{} \in \{ <, >, = \}.
\end{array}\]
The constructions include the standard first-order logical quantifiers, which range over the set of
flows $F$ being examined for the superflow hypothesis in question, the Boolean connectives, and
simple cardinality constraints over $F$. An example of such cardinality constraints would be the
constraint $|\{ \dstip(f) \mid f \in F \}| \geq 200$, indicating that we see at least 200
constituent flows as part of a superflow grouping.

\begin{example}[Chat session hypothesis]
One characteristic of a chat session between two hosts would be the exchange of back and forth
messages, each of which is smaller than the minimum transmission unit:
\begin{alignat}{1}
  h_{\text{chat}}(F) & = \forall f, f' in F, \nonumber \\
                     & \qquad (\srcip(f) = \srcip(f') \land \dstip(f) = \dstip(f') \lor \nonumber \\
                     & \qquad \srcip(f) = \dstip(f') \land \dstip(f) = \srcip(f')) \land \nonumber \\
                     & \qquad \numbytes(f) \leq 1500.
  \label{eq:sf:chat}
\end{alignat}
\end{example}

\begin{example}[Webpage fetch hypothesis]
Another example would be to delineate sessions involving a webpage fetch. A simple webpage fetch
may be modeled as a sequence of HTTP requests (heading either to TCP port 80 or port 443), each of
which closely follows a preceding DNS request (heading to UDP port 53):
\begin{alignat}{1}
  h_{\text{web}}(F) & = \forall f \in F, \dstport(f) \in \{ 80, 443, 53 \} \land \nonumber \\
                    & \quad (\dstport(f) \in \{ 80, 443 \} \implies \nonumber \\
                    & \quad \quad \exists f' \in F, 0 \leq \timestart(f) - \timestart(f) \leq 300 \text{ s} \land \nonumber \\
                    & \quad \quad \qquad \dstport(f') = 53).
  \label{eq:sf:web}
\end{alignat}
\end{example}


\subsection{Efficiently Decomposing Flow Streams}
\label{sub:sf:alg}

The central computational problem with superflow hypotheses is to identify a superflow
decomposition. The generality and nested quantifiers in the language of the previous section makes
this challenging. We will now identify some restrictions on superflow hypotheses that make the
decomposition problem tractable.

We start by focusing on the model-checking problem: Given a hypothesis $h$ and a set of flows $F$,
determine whether it is the case that $h(F) = \true$. We say that a hypothesis $h$ can be
\emph{efficiently monitored} if the model checking problem can be solved in a streaming manner in
time linear in $|F|$ and with memory that is independent of the size of the flow stream.

\begin{claim}[Efficient hypothesis monitoring]
Let $p(f_1, f_2)$ be a binary predicate which is either:
\begin{inparaenum}[(\itshape a\upshape)]
\item transitive (i.e., for all flows $f_1$, $f_2$, $f_3$, $p(f_1, f_2) \land p(f_2, f_3) \implies
  p(f_1, f_3)$), or
\item satisfies the property that $p(f_1, f_2) = \true \land p(f_1, f_3) = \true \implies
  p(f_2, f_3) = \true$.
\end{inparaenum}
In both cases, the hypothesis $h_{\text{trans}}(F) = \forall f_1, f_2 \in F, p(f_1, f_2)$ can be
efficiently monitored.
\end{claim}
This follows because upon seeing a new element $f_3$ of the flow stream, the monitoring algorithm
only needs to evaluate $p(f_1, f_3)$ for some arbitrarily pre-selected element $f_1$ of the flow
stream rather than comparing each pair of flow records previously encountered. As an example, both
binary constraints appearing in $h_{\text{scan}}$ are of this form. The constraint in
$h_{\text{chat}}$ is also transitive, so it follows that both hypotheses $h_{\text{scan}}$ and
$h_{\text{chat}}$ can be efficiently monitored.


We next turn our attention to the problem of constructing maximal decompositions. Observe that all
three hypotheses from Section~\ref{sub:sf:lang} lend themselves to a greedy superflow construction
procedure, by which one can repeatedly add new flow records to an existing candidate superflow until
it fails to satisfy the hypothesis in question. In particular, we have:
\begin{claim}[Superflow decomposition]
Let $h$ be an efficiently monitorable superflow hypothesis which is also \emph{subset closed}: i.e.,
whenever $h(F) = \true$, it is also the case that $h(F') = \true$, for all subsets $F' \subseteq
F$. In this case, a maximal superflow decomposition of a set of flows $F = \{ f_1, f_2, \dots,
f_n \}$ can be constructed in linear time, and with memory proportional to the number of reported
superflows.
\end{claim}

%% file: src/studies.tex
\section{Superflow-guided Data Reduction}
\label{sec:studies}
In this section, we discuss several case studies for a superflow construct and discuss the effectiveness of a superflow construct.  Recall from \S\ref{sec:sf} that a SuperFlow represents a {\em hypothesis} about a class of network traffic, and that the same individual flows may be present in an ensemble of superflows representing competing hypotheses. In order to argue for the efficiency of a superflow implementation, we must demonstrate that a superflow will increase EPAH (events per analyst-hour) processed. In the absence of an operations floor, we use on-disk footprint as a proxy for EPAH on the thesis that reducing the on-disk footprint reduces the query time for an associated phenomenon, and by reducing that query time, we increase the number of events an analyst can process.

In order to explore the superflow concept, we have created traffic traces on a large networking testbed using different client/server and service combinations in isolation. By creating clean traces focusing exclusively on specific classes of traffic, we can examine the construction and description of superflows and determine which attributes are necessary for effectively describing the superflow.  For this work, we have considered two scenarios: a website and scanning. 
The websites we examine in this study are multi-hosted, CDN-based and spread across multiple servers providing images, multimedia data, advertising, user tracking and JavaScript.  Modern webpages are often comprised of fetches from dozens or hundreds of different websites.  

Scanning, systematically targeting open ports on a network in order to determine the presence of vulnerable services, is an excellent target for a superflow formulation due to the disproportionate footprint scans leave in network traffic summaries.  Scans consist of a large number of packets with slightly different addresses, meaning that a small scan (256 hosts) may take up hundreds of more records than a long, multi-terabyte, data transfer.  

The remainder of this section is structured as follows, \S\ref{ss:footprints} discusses the {\em data footprints} of NetFlows, this discussion lays the groundwork for discussing how different SuperFlow representations will compare against equivalent NetFlow formulations and show how efficient the superflow has to be in order to substitute for a NetFlow.  \S\ref{s:ws} examines a modern website as a potential superflow construct; here we use data from web browsing sessions to show that a potential superflow is constructible and that it will substantially reduce the data footprint.  Finally \S\ref{s:scans} examines scanning data collected {\em in situ} from our darkspaces to show expected values for data reduction based on the types of scans seen.

\subsection{Estimating Data Footprints}
\label{ss:footprints}
\begin{figure}
    \includegraphics[width=3.5in]{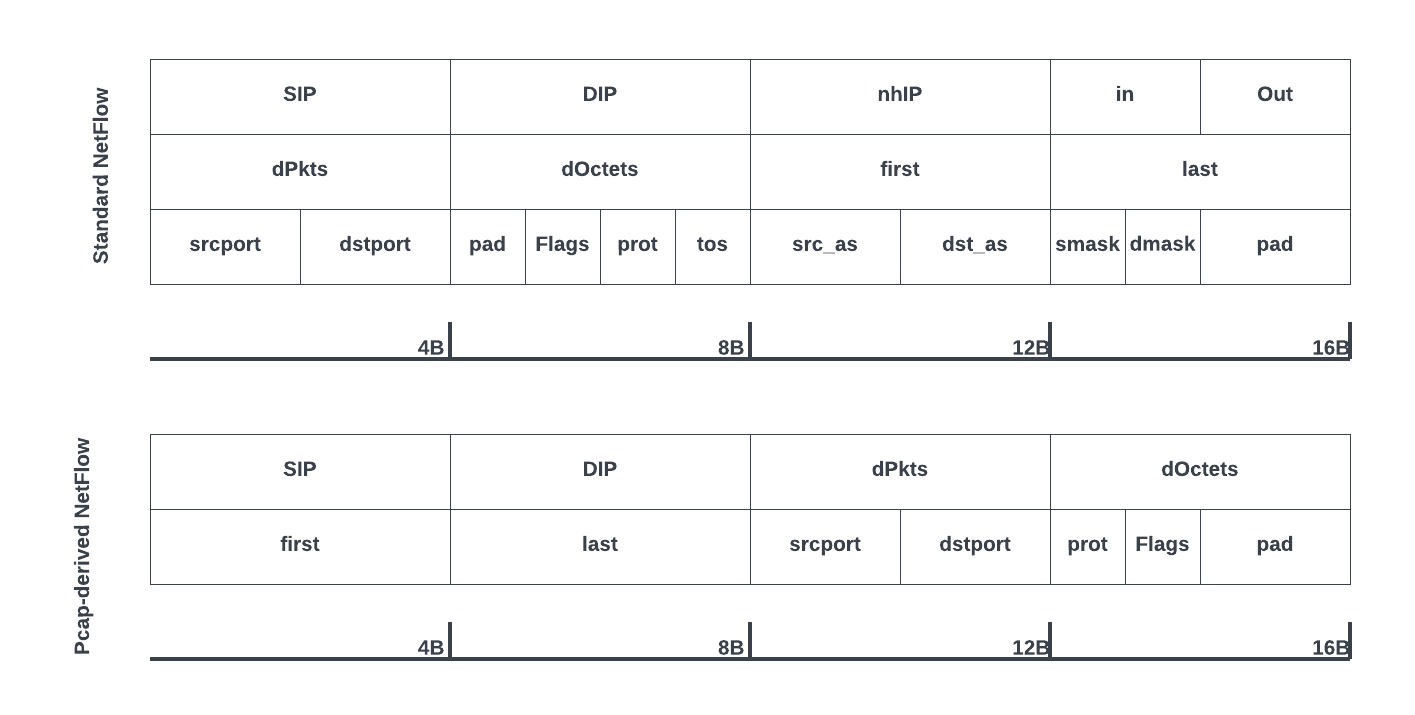}
    \label{f:basefoot}
    \caption{The Basic Footprints for NetFlow as collected by the router or through PCAP}
\end{figure}

NetFlow is a compact fixed-size representation of traffic data, which lends itself to random access and to representation in highly efficient data stores such as columnar databases. Figure~\ref{f:basefoot} shows this compact representation; each grid in this figure and the footprint diagrams that follows shows the footprint of netflows and our theoretical superflow constructs.  As this figure shows, standard V5 NetFlow as collected by the router has a 48-byte footprint.  However, as discussed in \S\ref{sec:intro} there are a number of pcap to flow tools which generate their own netflow representations; these representations can throw away router-generated data such as the next hop IP and the input or output interfaces.  This smaller, 32-byte footprint is what we will consider the default flow size for our data reductions. 

\subsection{Modern Webpage Analysis}
\label{s:ws}
\begin{figure}
    \centering
    \includegraphics[width=3in]{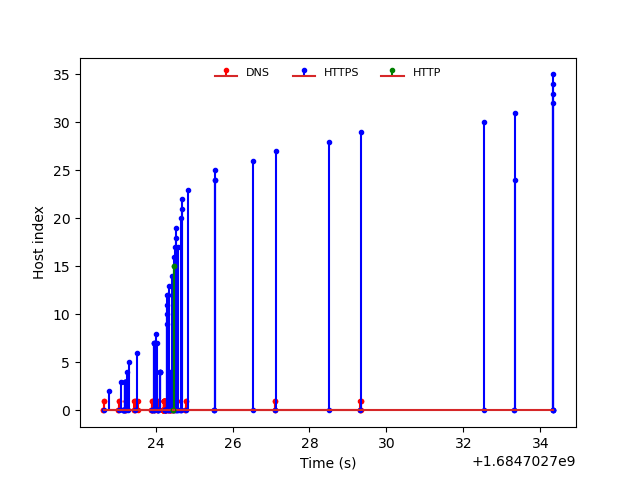}
    \caption{Discrete Sites Contacted to Fetch CNN's Homepage}
    \label{f:sitefetch}
\end{figure}

Modern webpages are assembled from dozens of files stored on discrete webservers; examples include the homepage of a news site like the New York times or a commerce site such as eBay.  These sites rely on the browser to fetch information from multiple discrete locations, resulting in multiple flows across different protocols (\ie, DNS, HTTP, HTTPS and QUIC, along with streaming video protocols) to produce a single page.  These interactions can become quite complex and large; for example, Figure~\ref{f:sitefetch} shows the number of discrete sites contacted by a browser when it fetches a page from CNN.  This fetch was constructed using a single browser fetching in private mode, limiting the potential for unrelated page fetches.  As Figure~\ref{f:sitefetch} shows, the browser contacts 36 different sites during the course of operations.  A single page fetch consists of 228 flows to 36  IP addresses.

\begin{figure}
    \includegraphics[width=3.5in]{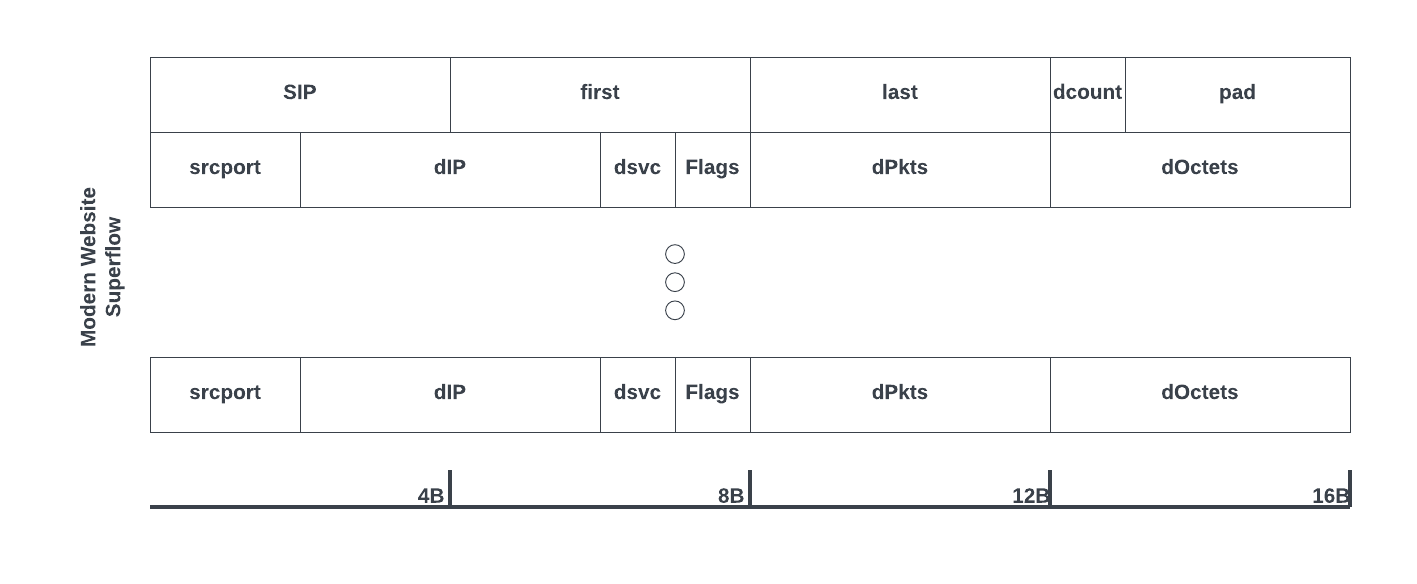}
    \caption{Footprints for Modern Website representation}
    \label{f:webfoot}
\end{figure}

Figure~\ref{f:webfoot} shows the footprint for a potential website superflow.  This superflow has a footprint of (16 + 16 $\cdot$ $\textrm{dcount}$) bytes, where $\textrm{dcount}$ is the number of subsidiary sites contacted during website construction.  This representation also merges destination port and service (which can include UDP/53 (DNS), UDP/443 (QUIC), TCP/443 (HTTPS), TCP/80 (HTTP) and then a variations such as TCP/8080) into a single byte value. 

The efficiency of a webpage superflow is driven by two factors: the number of sites comprising the webpage and the number of sites encoded in the superflow representation.  The CNN example, as noted, requires 36 addresses, which using the NetFlow format given in \S\ref{ss:footprints}, results in a total footprint of 1152 bytes, compared to 592 for the corresponding superflow formulation. Craigslist, by comparison, requires access to only 4 additional sites, yielding a 160 byte footprint using flows, 96 using superflows.  

\subsection{Scan Analysis}
\label{s:scans}
To examine the efficacy of superflow recognition and usage {\em in situ}, we collected and analyzed a set of candidate scan data from our institution's dark spaces.  A dark space is a collection of contiguous IP addresses which are routable, but do not have a responding host or DNS name.  Traffic to a dark space is suspicious because it was initiated by an outside organization due to a number of different phenomena, notably scanning, backscatter and misconfiguration.  Given our specific interest in scan summarization, we filtered the traffic to contain the most obvious scanning packets, these are TCP packets with a low ACK flag (indicating that the packet is not a response).  Out of the total traffic observed in any 24 hour sample period, this class of tcp traffic makes up 62\% of overall traffic (32\% of the remaining volume is any other TCP traffic, while ICMP, DNS and GRE make up remaining 6\%).

Using those packets, we developed an estimate for the potential impact between flows and a potential superflow.  To do so, we examine the number of flows we expect to reduce as a function of the likelihood of encountering a full 256-address scan.   The reduction is the expected number of flows within the dataset that would be replaced by a superflow.  For example, a superflow representing scans across a /24 would replace 256 flows. 

\begin{figure}
    \includegraphics[width=3.5in]{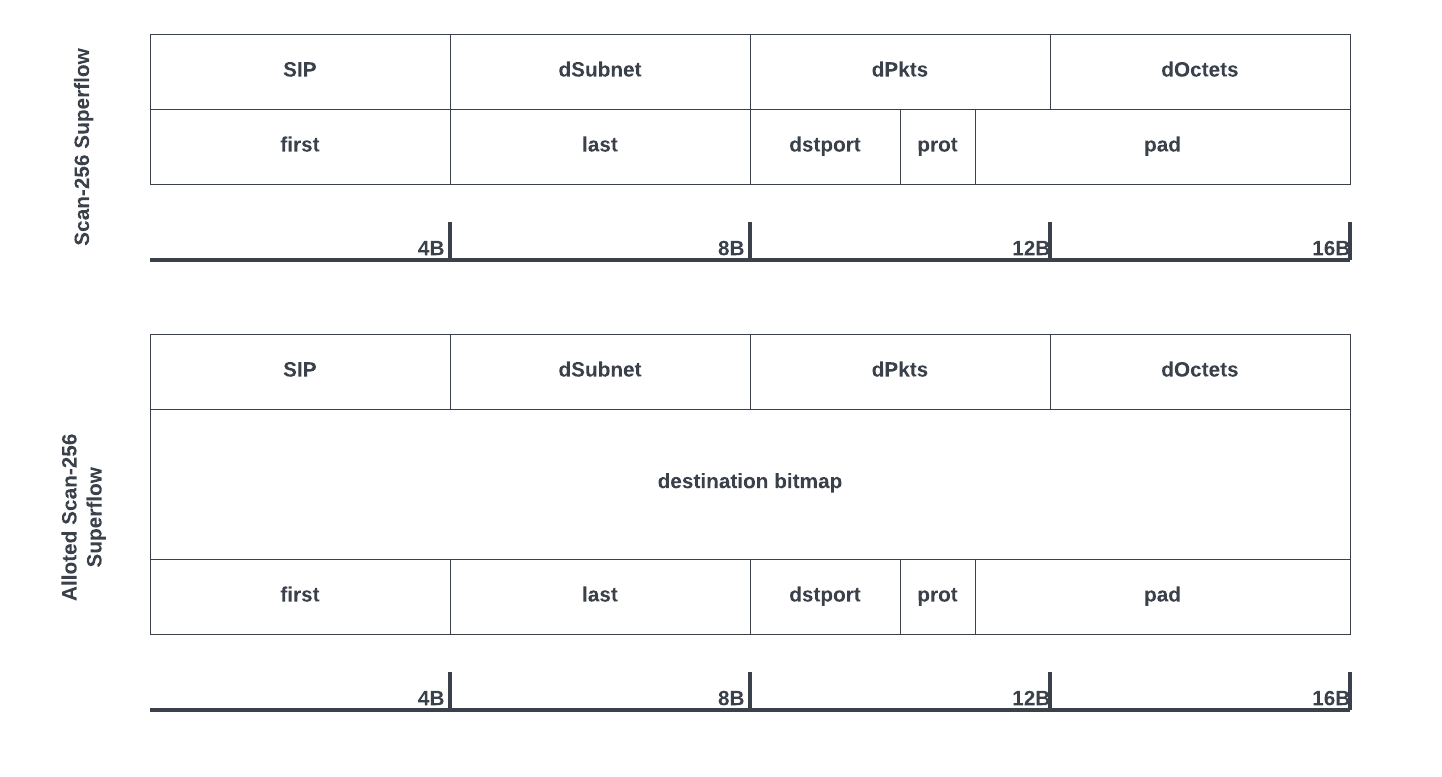}
    \caption{Footprints for Full and Allotted Scan-256}
    \label{f:scanfoot}
\end{figure}

Given this assumption, we define a {\em scan-256} superflow as a superflow which describes scanning between an individual host and a /24.  This superflow has a disk footprint shown in Figure~\ref{f:scanfoot} and is detected using the hypothesis in Equation~\ref{eq:sf:scan} with $c=256$.  We note in passing that this approach treats scan detection as a given; that is, the scan-256 superflow is identifying and isolating {\em obvious} scanners as opposed to differentiating very slow and subtle scanners.  While such scanners exist, identifying them becomes easier once the scan-256 flow has removed the obvious and noisy scanners from the analyst's workflow.

\begin{figure}[t]
\centering
\includegraphics[width=3in]{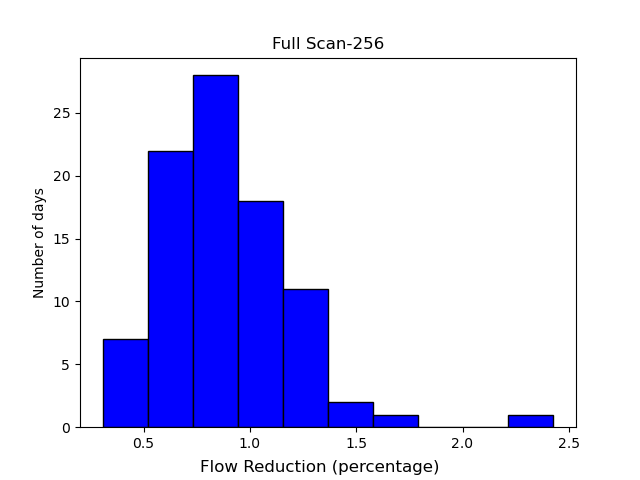}
\caption{Flow Reduction for Full Scan-256 Superflow}
\label{f:frf}
\end{figure}

\begin{figure}[t]
\centering
\includegraphics[width=3in]{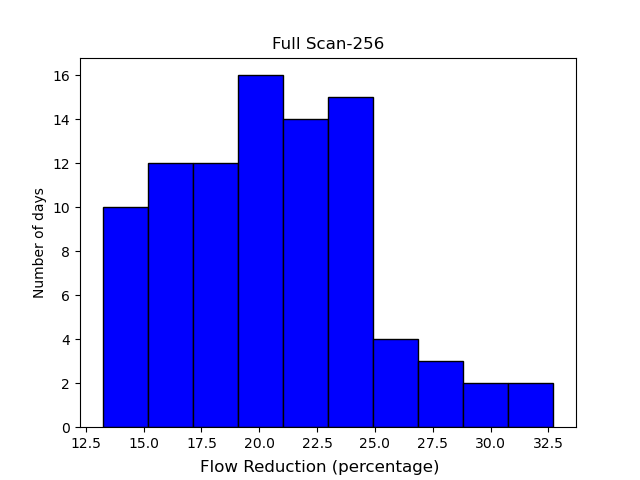}
\caption{Flow Reduction for Allotted Scan-256 Superflow}
\label{f:fra}
\end{figure}

Applying the initial scan-256 superflow rules to our darkspace data results in a small reduction, shown in Figure~\ref{f:frf}.  As this figure shows, replacing qualifying flows with scan-256 superflows reduces the total flow footprint by between 1/2 and 2.5\%.  This shows that while the on-disk reduction for a full 256 scan is substantial, there aren't enough of them to significantly reduce the total on-disk footprint.  Scanners rarely scan every address in a /24; often they will skip addresses such as x.255 or x.0.  To compensate for this, we considered an alternative structure we called an {\em allotted scan-256}.  The allotted scan-256 allots a table of IP addresses in order to indicate that the attacker skipped some subset of the total subnet; for the purposes of this paper we set the allotment to 32 IP addresses, creating a scan-256 superflow for flowsets with as little as 224 addresses or as many as 256, effectively using Equation~\ref{eq:sf:scan} with $c=224$.  The impact of this change is substantial; Figure~\ref{f:fra} shows the estimated flow footprint reduction for allotted scan-256 superflows, as this figure shows, the reduction is now between 12\% and 32\%.  

Figure~\ref{f:scanfoot} shows potential footprints for full and allotted scan-256 superflows.  As this figure shows, the full superflow is the same size once accounting for padding.  Given that a single /24 scan will comprise at least 256 flows, this is a substantial reduction.  Given that a full scan-256 will substitute for 256 flows, this results in a 32 byte footprint, as compared to the 8 kilobyte footprint for a full scan.



%% file: src/future.tex
\section{Discussion and Future Directions}
\label{sec:future}

We expect to implement superflows within the SiLK toolset by adding tools to construct the flows and then query them using the current {\tt rwcut} and {\tt rwfilter} tools. 

\paragraph{\textbf{Building new superflow class libraries.}}
In future work, we intend to develop a dictionary of different classes of superflows.  We expect that this dictionary will include the superflow classes already discussed, as well as chat protocols, email and peer to peer traffic.  Chat protocols, such as XMPP, Signal, and VOIP protocols, have distinct behaviors, notably jittery packets smaller than the MTU.  Other areas of interest include peer-to-peer protocols such as Bittorrent and SMTP.  Mail interactions are particularly important because, in addition to representing a significant fraction of Internet infrastructure, require collating information across at least 3 different protocols -- DNS, SMTP and POP3 or IMAP and optionally HTTP/S.  

\paragraph{\textbf{Vantage and confounders.}}
Also of import are the issues of \emph{vantage} and \emph{confounders} for superflow generation.  Vantage refers to the impact that sensor placement has on data collection; for example, modern websites as discussed in \S\ref{sec:studies} consist of multiple calls from a web client to multiple discrete servers, however modern interactive sites may also involve client requests from servers within the website to other servers, such as a database authenticating the user's identity.  This means that the flow data collected from a client's vantage may show different data than the flow data collected from one of the server's vantage. Tightly tied to the issue of vantage are the issue of {\em confounders}, these are middleboxes (such as NAT's) which affect assumptions about the identify of IP addresses across multiple flows.  Superflows must assume that some elements (such as client IP addresses) remain consistent and distinguishable. We intend to extend the superflow hypothesis language to express network topologies and to automatically reconcile data
from multiple sensors across the network.

\paragraph{\textbf{Expanding the scope of superflow constructs.}}
The current superflow formalism is based on relational logic~\cite{Alloy}, and we provide linear-time algorithms for many superflows
expressed in this formalism. As part of future work, we will expand the range of superflow hypotheses that can be
expressed and develop algorithms that can more efficiently decompose flow streams. We will also explore the possibility
of incorporating temporal patterns in superflow hypotheses and opportunities to automatically learn these relationships
using techniques such as Granger Causality~\cite{Granger}.

Another area of note is the ability to add post-processing data to superflows.  As noted in our discussion on the web superflow in \S\ref{sec:studies}, CDN's make up a significant amount of modern website traffic, and the round-robin DNS allocation used by many CDN's can result in multiple IP addresses which point to identical content servers.  

Finally, we need to further explore the need for superflows to describe alternative hypotheses within the superflow.  As noted in the Scan-256 example in \S\ref{sec:studies}, the allotted scan-256 provides more flexibility and summaries in exchange for a small initial storage overhead.  As the superflows are intended to improve operational response, including annotations about exceptional behavior (such as failed connections in a web superflow) can improve analyst efficiency at a small overhead cost. 